\documentclass[conference,10pt]{IEEEtran}
\usepackage{amsmath, amssymb}
\usepackage{tabularx} 
\usepackage{url}     
\usepackage{graphicx} 
\usepackage{lipsum}
\usepackage[T1]{fontenc}

\newcounter{protocol}
\newenvironment{protocol}[1]
  {\par\addvspace{\topsep}
   \noindent
   \tabularx{\linewidth}{@{} X @{}}
    \hline
    \refstepcounter{protocol}\textbf{Protocol \theprotocol:} #1 \\
    \hline}
  { \\
    \hline
   \endtabularx
   \par\addvspace{\topsep}}

\newcommand{\sbline}{\\[.5\normalbaselineskip]}

\PassOptionsToPackage{dvipsnames,table}{xcolor}

\usepackage{xcolor}

\PassOptionsToPackage{hyphens}{url}
\usepackage[
  colorlinks=true,
  linkcolor=purple,
  citecolor=purple,
  urlcolor=purple,
  pdfauthor={},
  pdftitle={},
  pdfsubject={},
  pdfkeywords={},
  bookmarks=false,
]{hyperref}

\hypersetup{colorlinks=true}

\usepackage{amsmath,amsfonts}
\usepackage{graphicx}
\usepackage{textcomp}

\usepackage[utf8]{inputenc}
\usepackage[english]{babel}

\usepackage[
  sorting=none,
  style=numeric-comp,
  backend=biber,
  isbn=false,
  url=false,
  doi=false,
  url=false,
  mincrossrefs=100,
  maxnames=12,
  giveninits
]{biblatex}

\AtEveryBibitem{\clearname{editor}}

\usepackage{caption}
\usepackage{subcaption}

\usepackage{pifont}
\usepackage{amsmath}
\usepackage{mathtools}
\usepackage{xspace}
\usepackage{mathtools}

\usepackage{flushend}

\usepackage{booktabs}
\usepackage{siunitx}
\usepackage[normalem]{ulem}
\usepackage{setspace}
\usepackage{multirow} 
\usepackage[commandnameprefix=ifneeded,truncate=hyphenate]{changes}

\usepackage[nopostdot,nonumberlist,acronyms]{glossaries}

\usepackage{csquotes}

\usepackage{algorithmic}
\usepackage{algorithm}

\usepackage{float}
\usepackage{graphicx}
\usepackage{cuted}
\usepackage{capt-of}

\usepackage{multirow}
\usepackage{diagbox}
\usepackage[table]{xcolor}

\usepackage[normalem]{ulem}
\usepackage{textcomp}
\usepackage{xcolor}
\usepackage{xspace}

\newcommand{\name}{\mbox{\textsc{Overdraft}}}

\IEEEoverridecommandlockouts
\begin{document}

\title{Taming Double-Spending in Offline Payments with Reputation-Weighted Loan Networks
\thanks{This work was funded by NWO/TKI grant BLOCK.2019.004.}
}


\author{
\IEEEauthorblockN{Nektarios Evangelou, Rowdy Chotkan, Bulat Nasrulin, J\'er\'emie Decouchant}
\IEEEauthorblockA{
    \textit{Delft University of Technology}\\
    Delft, The Netherlands \\
    }
}

\maketitle

\thispagestyle{plain} 
\pagestyle{plain}

\begin{abstract}
Blockchain solutions typically assume a synchronous network to ensure consistency and achieve consensus. 
In contrast, offline transaction systems aim to enable users to agree on and execute transactions without assuming bounded communication delays when interacting with the blockchain. Most existing offline payment schemes depend on trusted hardware wallets that are assumed to be secure and tamper-proof. While this work introduces \name{}, a novel offline payment system that shifts the reliance from hardware to users themselves. \name{} allows potential payment receivers to assess the likelihood of being paid, allowing them to accept transactions with confidence or deny them. 
\name{} achieves this by maintaining a loan network that is weighted by online reputation. This loan network contains time-limited agreements where users pledge to cover another user's payment if necessary. For example, when a payer lacks sufficient funds at the moment of commitment. Offline users rely on the last known view of the loan network---which they had access to when last online---to determine whether to participate in an offline transaction. This view is used to estimate the probability of eventual payment, possibly using multiple loans. Once online again, users commit their transactions to the blockchain with any conflicts being resolved deterministically.  
\name{} incorporates incentives for users and is designed to be resilient against Sybil attacks.  
As a proof of concept, we implemented \name{} as an Ethereum Solidity smart contract and deployed it on the Sepolia testnet to evaluate its performance. 
\end{abstract}


\section{Introduction}
\label{sec:introduction}

The increasing dependence on Internet connectivity for financial transactions overlooks scenarios where an affordable and reliable connection cannot be assumed~\cite{fund2016international,liu2021stay,buteau2021emerging}. This technological gap highlights the need for robust offline payment solutions to address the shortcomings where traditional Internet-based systems are ineffective~\cite{poon2016bitcoin,park2017opera,baqerAMPS17,lindNEKPS18,blokzijl2021eurotoken,gupta2022m,chu2022review}. While traditional payment methods such as cash, checks, postal orders, and bank transfers remain practical, they suffer from security, speed, and scalability constraints, making them inadequate for the demands of the modern, digital economy where high availability is critical~\cite{bezovski2016future,de2018societal,brunnermeier2019digitalization}.

The absence of dependable offline payment mechanisms hinders financial inclusivity and exposes businesses and individuals to operational risks during network congestion or failures~\cite{una2023fintech}. These limitations are most noticeable in remote regions where the technological gap is most prominent~\cite{anakpo2023policies}. Offline payment systems can leverage trusted components to address the issue of double-spending when Internet connectivity is unavailable~\cite{lindNEKPS18,Christodorescu2012}. However, the reliance on trusted components introduces single points of failure and trust. 

In this work, we leverage the concept of loan networks, a mechanism that introduces a trust-based system allowing for secure and scalable transactions without continuous online verification~\cite{ChengWZ020,ChengNZ23}. The core of our approach lies in a reputation-weighted loan network, enabling network entities to guarantee (loan) tokens for one another. This loan mechanism enables a self-regulating ecosystem where trust plays a vital role as an indicator and validator of the trustworthiness of nodes in the network~\cite{resnick2000reputation}.

Our research focuses on developing a framework to mitigate the risk of double-spending, a well-known concern in offline transactions~\cite{everaere2010double,karame2012double}, using a probabilistic evaluation based on trust scores and mutual loans. We construct a theoretical framework that employs the Web of Trust model~\cite{zimmermann1995official} with the practicalities of offline transactions. We devise an algorithm that traverses a node's loan network to find a loaning node to cover the offline transaction amount when the node cannot pay. Intuitively, loans are valid for a pre-specified duration when published on-chain, and offline users can rely on the loans they know or learn about to accept (or deny) a payment. More specifically, a payer accepts a payment if it considers that it will (resp. will not) be online on time to claim its payment and is confident that the payer's loan network will execute the payment.
Furthermore, we explore the potential of predicting double-spending incidents in offline settings, laying the groundwork for a protocol that ensures transaction integrity, authenticity, and non-repudiation without real-time internet access.

\subsection*{Contributions.}
\begin{itemize}
    \item We design \name{}, the first offline payment framework that mitigates double-spending risks without relying on trusted hardware or online interactions. To do so, \name{} leverages a reputation-weighted loan network that allows a payment receiver to evaluate the guarantees a payer has to offer through its loan network.  
    \item We implement the online part of \name{} using a Solidity Ethereum smart contract that maintains the list of active loans, and its offline part in Python. \name{} requires standard asymmetric cryptography. 
    \item We provide incentives for users to participate in loan networks and demonstrate that Sybils cannot profit from \name{}.   
    \item We evaluate the performance, resource consumption, and costs that using \name{} incurs.  
\end{itemize}

\noindent This paper is organized as follows. 
\S\ref{sec:related_work} surveys the related work.
\S\ref{sec:system_overview} provides a high-level overview of \name{}.
\S\ref{sec:system_details} describes \name{}'s system details.
\S\ref{sec:sybil-tolerance} explains how \name{} achieves Sybil tolerance.
\S\ref{sec:performance_evaluation} presents our performance evaluation, which discusses \name{}'s resource consumption, throughput, latency and fees.  
\S\ref{sec:conclusion} concludes this paper and discusses limitations and possible extensions. 

\section{Related Work}
\label{sec:related_work}

\hspace{\parindent} \textbf{E-cash systems.} Okamoto and Ohta~\cite{okamoto1991universal} listed the six key properties of an ideal e-cash system~\cite{chaum1983blind}: independence, security, privacy, divisibility, transferability, and offline payment. 
E-cash systems either prevent double spending or detect it. Double spending prevention requires assuming a trusted hardware or software~\cite{boly1994esprit}, or a trusted third party~\cite{brands1994untraceable,everaere2010double}, while its detection can be achieved with online distributed audits~\cite{yacobi1999risk} or when a coin is claimed twice online.  

\textbf{Layer-one transactions.} A first way to execute user transactions using a blockchain is to have them disseminated to the whole network and ordered by its consensus service~\cite{nakamoto2008peer,buterin2014next}, which prevents double-spending under the assumption that the blockchain is secure. However, submitting a transaction this way requires network access, and therefore prohibits offline payments, and has been getting increasingly expensive. 
 
\textbf{Layer-two transactions.} Layer-two protocols~\cite{gudgeon2020sok} run on top of layer-one blockchains and eliminate the need to broadcast every transaction across the entire network. Instead, they allow users to exchange authenticated transactions off-chain. Examples of such protocols include payment channel networks~\cite{poon2016bitcoin}, virtual channels~\cite{dziembowski2019perun}, anonymous payment channels~\cite{lindNEKPS18} and rollups. The vast majority of layer-two protocols require their users to be regularly online to participate in possible on-chain disputes. A notable exception is Teechain~\cite{green2017}, a payment network that removes this assumption by relying on trusted execution environments to support offline payments.

\textbf{Offline payment systems.} To prevent double spending and implement user wallets, offline payment systems have mostly been relying on trusted hardware. UEPS~\cite{anderson1992ueps} is a payment system that relies on offline payment terminals and trusted smartcards. BlueWallet~\cite{bamert2014bluewallet} is a proof-of-concept hardware wallet that can sign and authorize transactions and relies on Bluetooth Low Energy for communication. DigiTally~\cite{baqerAMPS17} is an offline mobile payment system that uses a trusted overlay SIM card. 

\textbf{Offline blockchain payments and CBDCs.} PureWallet~\cite{igboanusi2021blockchain} is a blockchain system that supports offline transactions using near-field communication (NFC) and a tamper-resistant NFC Secure Element (SE). Central banks around the world have been considering issuing Central Bank Digital Currencies (CBDC). CBDCs either rely on the online settings, like Platypus~\cite{wust2022platypus} and Peredi~\cite{kiayias2022peredi}, or assume a trusted hardware~\cite{christodorescu2020towards}.    

Contrary to previous works, our offline payment system, \name{}, does not assume a trusted component or a synchronous network access to handle possible on-chain disputes. As a consequence, \name{} also does not aim to completely eliminate double-spending in offline payments. Instead, it mitigates double-spending risks by allowing users to evaluate the probability that they will be paid, either directly or through a loan network, and independently decide to accept or decline an offline payment.       

\section{Protocol Overview}
\label{sec:system_overview}

\begin{figure*}[t]
    \centering
    \includegraphics[scale=0.7]{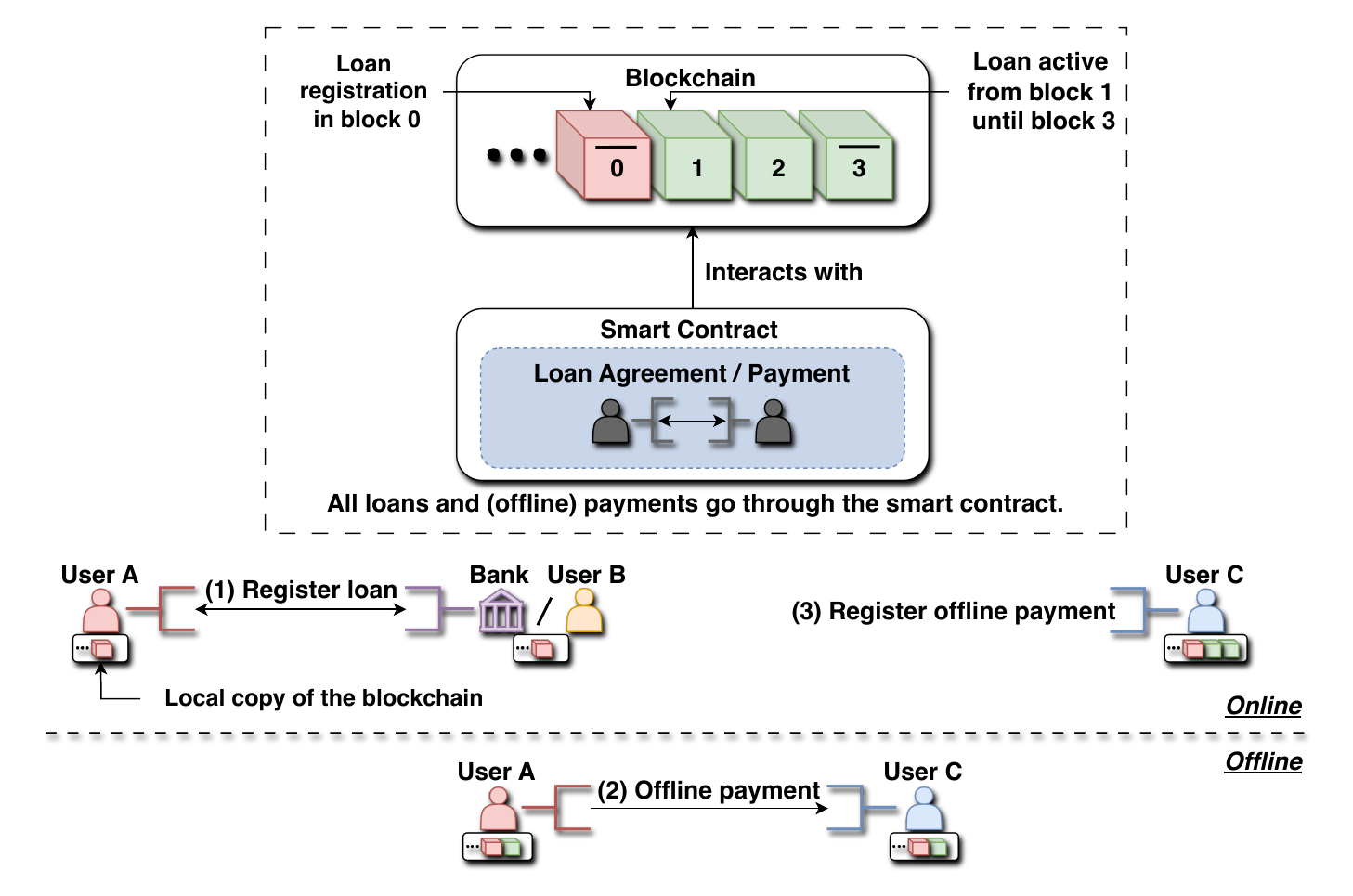}
    \caption{Overview of \name{}}
    \label{fig:system-architecture}
\end{figure*}

Figure~\ref{fig:system-architecture} illustrates \name{}'s architecture and the interactions between its components and users in a typical set of transactions. 

We introduce the successive steps involved in an offline payment in the following. 

\textbf{Loans.} \name{} leverages loans that users directly contract among themselves. While loans can be made arbitrarily complex, we consider that they specify a lender $\mathcal{A}$ and a borrower $\mathcal{B}$, an amount $L \in \mathbb{R}^+$ that the lender agrees to pay in place of the borrower if needed, a starting time $T_o \in \mathbb{T}$ (i.e., a block height), a time validity $T_d \in \mathbb{T}$ (counted in numbers of blocks) and a maintenance fee $F_o \in \mathbb{R}^+$ that the borrower regularly has to pay while the loan is active. A loan is digitally signed by all parties it involves. 

\textbf{System model.} \name{} relies on a blockchain that supports smart contracts. We assume that the blockchain is secure. Users participate in an offline payment outside of \name{}'s protocol, without relying on a synchronous network, and have them processed by \name{}'s smart contract when they are back online. The amount of time users can stay offline is not bounded. However, in practice, users would estimate the time at which they would be back online to leverage the active loans they know of. 

\textbf{Threat model.} Users might misbehave and attempt to double spend or participate in transactions without having sufficient funds. 
\name{} relies on an online reputation scheme~\cite{nasrulin2022meritrank} that attributes a reputation score to each user (cf. \S\ref{sec:reputation}) to estimate the probability that a user misbehaves in a payment, e.g., a reputation of $r_i=0.5$ indicates that there is a $50\%$ chance that a payment of node $i$ will be unsuccessful. We consider that user reputation is maintained by a decentralized reputation system, such as MeritRank~\cite{nasrulin2022meritrank}, that avoids storing reputation-related information on the blockchain for cost reasons.

\textbf{Registering loans.} Users establish loan agreements among themselves, possibly while they are offline (step 1, Fig.~\ref{fig:system-architecture}), and then register them on the blockchain using \name{}'s smart contract (cf. \S\ref{sec:smartcontract}).  Note that if a node agrees to possibly lend a given amount in a contract, then the corresponding amount is immediately locked on its account until the loan expires to prevent Sybil attacks (cf. \S\ref{sec:sybil-tolerance}). For example, in Fig.~\ref{fig:system-architecture}, users $\mathcal{A}$ and $\mathcal{B}$ agree on a loan and submit it to \name{}'s smart contract, which inserts it in the blockchain in block $b_0$. This loan is active from blocks $b_1$ to $b_3$, which means that, under specific conditions, it might be used by an offline payment that is inserted in those blocks. 

\textbf{Taking reputation and loans offline.} While they are online, users can monitor the blockchain to maintain a local list of loans that \name{} has committed to the blockchain, and the reputation of all users. 
When they happen to be offline and want to participate in a payment, users can consult their local copy of the loans, which might be incomplete but only contains valid loans, and the reputation of other nodes. Users rely on the view of the reputation network they obtained while online. 

\textbf{Offline payments.} At first, users agree on and execute offline payments directly among themselves (step 2, Fig.~\ref{fig:system-architecture}) without synchronous communication with the blockchain and without involving trusted components. Users rely on their local reputation and loan network knowledge to decide for themselves whether they should participate in an offline transaction. To guide this decision process, the reputation of a user, either the payer or a lender, is used to approximate the probability that they would eventually pay a given amount online if needed (cf. \S\ref{sec:confidence}). Once users have agreed to participate in an offline payment, they can execute it (e.g., exchange some goods) and keep a signed transaction.

\textbf{Online execution of offline payments.} Whenever they are back online, users can send the offline transactions they participated in, either as payer or payee, to \name{}'s smart contract (step 3, Fig.~\ref{fig:system-architecture}; details in \S\ref{sec:smartcontract}). To execute a transaction, \name{} attempts to transfer the transaction's amount $A_t$ from the payer to the payee. If the payer's balance is insufficient, then \name{} will attempt to use its loan network to obtain funds from the nodes that vouched for the payer (possibly transitively). Different algorithms can be used to split the missing payment across all vouching nodes (e.g., limiting the depth at which vouching nodes might have to pay). The online execution of an offline payment is used to increase or decrease the reputation of the parties involved in it.    

\section{\name{} Design}
\label{sec:system_details}

This section discusses the main components of \name{} in detail: the reputation system, the smart contract that maintains loans and processes offline payments, and the algorithm that allows payee(s) to evaluate their confidence in an offline payment.

\subsection{Integrating reputation}
\label{sec:reputation}

MeritRank~\cite{nasrulin2022meritrank} is a decentralized and Sybil-tolerant reputation system where peers within a network actively observe and evaluate each other's contributions, which are recorded in a personal ledger. MeritRank generates a directed feedback graph, where vertices are peers, and weighted edges are the feedback a peer accumulates over time concerning another peer. Reputation scores are attributed to all nodes based on aggregated feedback, utilizing epochs to capture the dynamic nature of contributions.

\name{} employs a practical approach to calculating node reputation by integrating MeritRank's methodology with a decentralized ledger. This combination evaluates both the quantity and quality of node interactions, where the ledger tracks each node's transactions, both successful and unsuccessful. By focusing on failed transactions, our model determines each node's risk to the network. As a consequence, reputation scores are based on transaction volume and reflect nodes' reliability and trustworthiness, accommodating failures beyond their control and providing a transparent, secure basis for trust assessment within the network.

\subsection{Confidence in offline payments}
\label{sec:confidence}

Users who consider being the recipients of offline transactions rely on their view of the loan network to decide whether to accept or decline to participate in them. Each edge in the loan network indicates a contracted loan, and it connects a lending user to a potential recipient with a weighted edge. The reputation of a user indicates the probability that it would pay a promised amount (payment or loan). Based on the loan network, the potential offline payment recipient computes the distribution of the amount it might receive should one or more users be insolvent when eventually trying to get paid online. This procedure effectively simulates the processing of an offline transaction by \name{}'s smart contract, which might involve a dispute resolution using the loan network.

\begin{figure}[t]
    \centering
    \includegraphics[scale=0.6]{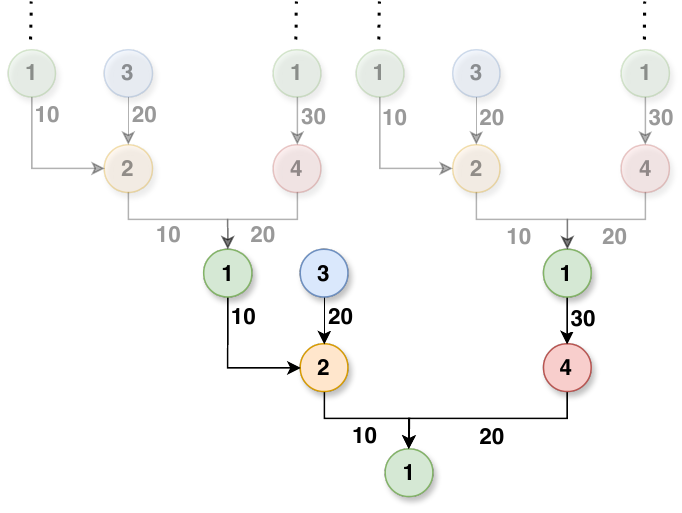}
    \caption{A simplified loan network that can be used to simulate the online processing of an offline payment by user 1. A cycle, such as the one created by users 1 and 2 loaning 10 coins to each other, might lead to infinite recursion during the randomized exploration if edges could be visited multiple times.}
    \label{fig:cycleexample}
\end{figure}

The confidence evaluation algorithm averages the result of several depth-first explorations of the loan network, starting from the payer vertex and following randomized edges to simulate users being insolvent. The complexity of this algorithm is practical, contrary to the exponential complexity of the brute-force algorithm that would evaluate all combinations of solvent and insolvent users. During a random exploration, an edge (i.e., a loan) can only be crossed once, but nodes can be visited multiple times. Figure~\ref{fig:cycleexample} shows what would happen when cycles are encountered in a random walk without prohibiting going through edges multiple times. We prevent infinite loops caused by cycles by tracking visited edges, which ensures the algorithm's termination. A random walk is halted when the amount it generates meets or exceeds the target transaction amount, and the running time is also controlled using a maximum exploration depth. 

\begin{algorithm}
\caption{Simulation of online dispute resolution using a loan network}
\begin{algorithmic}[1]
\REQUIRE $node$ (Node), $loaned\_amount$ (integer), $visited\_edges$ (set of Edges), $path$ (list of integer), $transaction\_amount$ (integer), $decay$ (float), $root\_node$ (Node), $max\_distance$ (integer)
\ENSURE $amount$ (integer)
\STATE $current\_path$ $\leftarrow$ $path + [node.node\_id]$
\IF{$node \text{ is } root\_node$ AND $\text{len}(current\_path) > 1$}
    \RETURN $0$
\ENDIF
\STATE $distance \leftarrow \text{len}(current\_path) - 1$

\STATE $decayed\_reputation {\leftarrow} node\_reputation \cdot {decay}^{distance}$
\STATE $will\_pay {\leftarrow}$ rand. boolean based on node's decayed reputation

\IF{$will\_pay$ is True}
    \RETURN $loaned\_amount$
\ELSE
    \STATE $amount \leftarrow 0$
    \STATE $edges$ $\leftarrow$ list of edges from $node.edges$ not in $visited\_edges$
    \FOR{each $edge$ in $edges$}
        \IF{$amount \geq transaction\_amount$ OR $distance \geq max\_distance$}
            \STATE break
        \ENDIF
        \STATE $visited\_edges.\text{add}(edge)$
        \STATE $predecessor \leftarrow edge.from\_node$
        \IF{$predecessor.id \text{ is } root\_node.id$}
            \STATE break
        \ENDIF
        
        \STATE $amount {\leftarrow} amount + \text{random\_walk}(predecessor$, $edge.loaned\_amount, visited\_edges$, $current\_path$, $root\_node)$
    \ENDFOR
    \RETURN $amount$
\ENDIF
\end{algorithmic}
\label{walkdecaypseudocode}
\end{algorithm}

Algorithm~\ref{walkdecaypseudocode} provides the confidence evaluation pseudocode. It starts from the payer's vertex and determines whether it would pay the transaction amount based on its reputation, which is used as a probability to be solvent. If the payer pays, then the algorithm returns the payment amount, otherwise, the algorithms will continue and study whether the users that agreed to loan money to the payer could pay for them. The algorithm contains two practical modifications. We incorporate a decay factor $D \in [0, 1]$ to model decreasing node influence with distance, inspired by trust dynamics in social and financial networks, ensuring distant nodes have less impact on the outcome~\cite{katona2011network}.
Furthermore, we impose a maximum distance $H$ to prioritize closer, more reliable nodes for endorsement, effectively addressing trust dilution over distance. 
The algorithm excludes visited edges to avoid cycles and iterates over a node's predecessors to avoid self-references. It accumulates loan amounts through recursion until all loans are exhausted. It terminates when the collected amount meets the amount of the offline transaction or when the maximum allowed distance to the payer is reached. 

During the execution of our algorithm, the amount $A(i)$ that a user $i$ would contribute towards the offline transaction is evaluated based on Equation~\ref{eq:recursive}.

\begin{equation}
A(i) = 
\begin{cases}
L_{i}, \text{with prob. $R(i) \cdot D^{H}$} \\
    \mathrm{min}(L_i, \sum_{j \in P(i)} A(N_{j}, V(E(j,i)))_{E(j,i) \notin S}), \\
    \hspace{1cm} \text{with prob. $1 - (R(i) \cdot D^{H})$}
\end{cases}
\label{eq:recursive}
\end{equation}

In this equation, $L_i$ is the amount user $i$ agreed to pay or loaned following the current edge that the algorithm considers, $R(i) \in [0, 1]$ is the reputation of user $i$; $P(i)$ denotes the vertices of the nodes that loaned some amount to user $i$; the edge from users $j$ to $i$ is a loan denoted as $E(j, i)$ and its value is $V(E(j, i))$, and $S$ is the set of previously visited edges. 

\subsection{Smart contract functionalities}
\label{sec:smartcontract}

\name{} uses a smart contract to register or close loan agreements and to process offline payments. Using a single smart contract distributes the fees between several parties by bundling operations. The loan agreement is encoded using the fields detailed in Table~\ref{table:contract_fields}.

\begin{table*}[!ht]
\caption{Loan Agreement fields with sizes and descriptions.}
\label{tab:loanfields}
\centering
\begin{tabularx}{\textwidth}{|l|l|X|}
\hline
\textbf{Field} & \textbf{Size (bits)} & \textbf{Description} \\
\hline
Agreement ID & 256 & The unique identifier of the loan agreement. \\
\hline
Public keys & 160 $\cdot$ 2 & The public keys of both parties. \\
\hline
Reputation scores & 256 $\cdot$ 2 & The reputation scores of both parties. \\
\hline
Loaned amount & 256 & The tokens of trust and the liability the loaning node accepts. This is used to verify whether it is feasible to loan the required amount. \\
\hline
Repayment time & 256 & The terms include the duration of blocks for repayment of the loaned amount, including the interest rate. \\
\hline
Dispute resolution & 256 & The mechanism to resolve disputes in the agreement for scenarios where there is a disagreement over agreement terms or the fulfillment. \\
\hline
Agreement duration & 256 & The amount of time an agreement will be usable. \\
\hline
Min. Open time & 256 & The minimum opening time when the agreement can be established. This value is required for the interest rate and the agreement between the parties. \\
\hline
Close time & 256 & The time when the agreement has been terminated. This value is required to calculate the final interest rate and end the agreement between the parties. \\
\hline
Opening fee & 256 & The fee to set up the agreement between two parties. Both parties need to pay the amount. \\
\hline
Closing fee & 256 & The fee to terminate the agreement between two parties. If the agreement is terminated and used in an offline payment, the recipient pays the fee. \\
\hline
Opening block & 256 & Represents the block number of the agreement opening transaction. \\
\hline
Closing block & 256 & Represents the block number of the agreement termination transaction. \\
\hline
Active & 256 & Represents whether the loan agreement is still active. If active, the amount has yet to be claimed and accrues interest rate. If inactive, the amount has been used, and the repayment time will start. \\
\hline
\end{tabularx}
\label{table:contract_fields}
\end{table*}

\textbf{Opening a loan.}
When \name{}'s smart contract receives a loan agreement, it verifies its validity and asks for it to be inserted in a block at a specified opening time or otherwise canceled. A loan specifies a validity period through start and end blocks, making the agreement enforceable only within the time frame that separates them. Once published on the blockchain, users can consider a loan in their offline confidence calculations. 
Activating a loan triggers a token-locking mechanism that locks the amount of the loan and the associated interests (see Sec.~\ref{sec:incentivization_model}) on the loaner's account, which is essential to prevent malicious users from inserting loans that are not backed by actual funds (see Sec.~\ref{sec:sybil-tolerance}). If the loaner does not have enough funds, then the loan is canceled, and it does not appear on the blockchain. 

\textbf{Closing a loan.} A loan is automatically closed after its end block. It is not possible to close a loan earlier because offline users might still consider it for their confidence computations. 

\textbf{Processing offline transactions}. Upon regaining online connectivity, nodes can reconcile offline transactions in which they participated with the blockchain to update their balance. The smart contract reviews loan agreements to settle balances based on these transactions, addressing offline payments and compensating the receiving nodes through a random selection of loaning individuals in case of dispute. This random selection utilizes a hash function that combines the current block timestamp, the previous block's \texttt{RANDAO} beacon value, and the user's public key address. Furthermore, the extra fees required by the smart contract traversing the loan agreements must be paid by the insolvent party and its loan network. Whether or not an offline payment is successfully executed without relying on the loan network respectively increases or reduces the reputation of a payer. 

For completeness, we provide pseudocode implementations of the smart contract routines for opening, closing, and processing loan agreements in Appendix~\ref{app:routines}. These routines perform standard contract logic and are omitted from the main text for brevity. A visual example illustrating how these routines interact in a full offline payment scenario is provided in Appendix~\ref{app:paymentprotocol}.

\subsection{Incentivizing loans}
\label{sec:incentivization_model}

\name{}'s loans are structured similarly to bank loans, i.e., the user benefiting from a loan must repay the amount with interest. The interest rate for a loan agreement is determined by several factors, including the loan amount, the duration of the loan, and the loaning node's reputation. We use the following interest rate formula to encourage participation, discourage malicious behavior, and fairly compensate for risks:

\begin{equation}
	I = \mathrm{max} \left( 0, \left( \alpha^{\beta} \cdot \frac{1}{1 + e^{-\zeta(R - R_{0})}} \right) + \left( \frac{\gamma}{365} \cdot \delta \right) \right)
\end{equation}

The total to be paid interest, $I$, is determined by a complex formula that considers several parameters: (1) the loan amount, $\alpha$; (2) the loaned percentage rate, $\beta$, which adjusts the interest based on the size of the loan; (3) the loan duration, $\gamma$ and (4) the annual percentage rate, $\delta$. 
The lending node's reputation, $R$, also plays a crucial role. Due to the characteristics of the sigmoid function, higher reputations result in increased interest rates. This effect is moderated by a penalizing midpoint constant, $R_0$, which serves as a threshold for reputation. The interest rate is adjusted based on how the lender's reputation compares to this midpoint. The steepness factor of the sigmoid, $\zeta$, further determines the rate's sensitivity to changes in reputation. While a loan is active, fixed interests are paid after every blockchain block.  

\begin{figure*}[t]
    \centering
    \begin{subfigure}{\columnwidth} 
       \centering
        \includegraphics[width=.7\columnwidth]{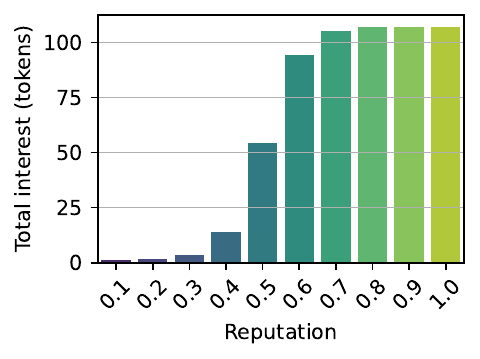}
        \caption{Total interest earned depending on reputation.}
        \label{fig:interest_rate_vs_reputation}
    \end{subfigure}%
    \hfill
    \begin{subfigure}{\columnwidth} 
        \centering
        \includegraphics[width=.7\columnwidth]{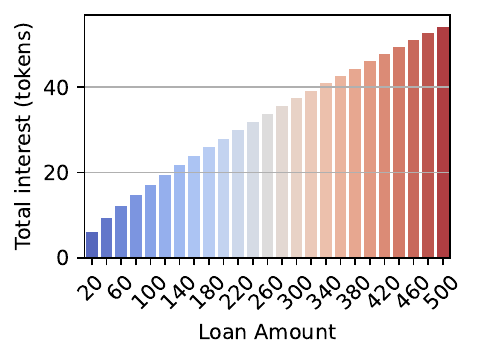}
        \caption{Total interest earned depending on the loan amount.}
        \label{fig:interest_rate_vs_loan_amount}
    \end{subfigure}
    \caption{Total interest earned depending on reputation and loan amount.}
    \label{fig:interest_rates_comparison}
\end{figure*}

The impact of reputation and the amount loaned on the interest rate is illustrated in Figure~\ref{fig:interest_rates_comparison}. The left bar chart (Fig.~\ref{fig:interest_rate_vs_reputation}) shows how interest rates vary with reputation, highlighting the benefits of maintaining a high reputation. We use the following values for the parameters in the interest rate formula: $\alpha=500$ tokens, $\beta=0.75$, $\gamma=100$ days, $\delta=5\%$, $R=0.5$, $R_0=0.5$, $\zeta=20$. The right bar chart (Fig.~\ref{fig:interest_rate_vs_loan_amount}) demonstrates the interest rates for different loaned amounts, underlining the incentive for nodes to engage in the lending process. 

\section{\name{} Analysis}
\label{sec:sybil-tolerance}

\subsection{Decay Strategies for Sybil-Tolerant Reputation}

\name{} uses MeritRank~\cite{nasrulin2022meritrank}, a Sybil-tolerant reputation mechanism, to maintain the reputation scores of users. MeritRank enhances Sybil tolerance through the use of three thresholds. First, the parallel report bound addresses vulnerabilities to parallel and cycle attacks,
limiting the total reputation across all involved Sybil nodes to at most the one of the first introduced Sybil node. Second, the serial report bound prevents indefinite reputation inflation from serial attacks by capping the total reputation gain from sequentially added Sybil nodes. Finally, the transitivity bound ensures that reputation propagation does not exceed the minimum reputation among all nodes on a given path, thereby preventing artificial reputation amplification and reflecting genuine node trustworthiness.

MeritRank implements decay strategies to enhance Sybil tolerance in networks through three decay parameters $\alpha$, $\beta$, and $\gamma$. The decay parameter $\alpha$ limits the propagation of reputation scores, curbing the influence of fake node chains by reducing the length of influence paths. The $\beta$ decay parameter counters structural vulnerabilities by penalizing nodes connected through bridge connections for forming separated components in the network, identified by a significant flow of random walks through a cut vertex. Finally, the $\gamma$ decay parameter combats reusing previous connection exploitation by diminishing the benefits of old attack edges, requiring continuous effort for reputation upkeep. These strategies include transitivity limitation, connectivity penalties, and epoch-based adjustments to strengthen network security.

\subsection{Splitting reputation over Sybils}

Let $R \in [0,1]$ be the reputation of a single node capable of lending amount $X$. Consider $\epsilon$ as a small fixed constant representing the penalty factor for splitting the reputation into multiple nodes. Let $R_{1}$ and $R_{2}$ represent two Sybil nodes such that each node's reputation is bounded by $R_{1}, R_{2} \in [0, R - \epsilon]$ and such that their combined reputation is slightly less than $R$, i.e., $R_{1} + R_{2} < R + \epsilon$~\cite{nasrulin2022meritrank}. 

The influence of the original node is $R \cdot X$.
Let $Y$ and $X - Y$ be the amounts of these Sybil nodes' loans, respectively, ensuring that the total amount remains $X$. 
The combined influence of the Sybil nodes is $R_{1} \cdot Y + R_{2} \cdot (X - Y)$. 
The reputation system penalizes splitting by reducing the effective reputation of each split node by at least $\epsilon$ for each split transaction, which gives 
$R \cdot X > (R_{1} + R_{2} - \epsilon) \cdot X$.
By writing $X = Y + (X-Y)$, this inequality can be rewritten as:
\[ R \cdot X > R_{1} \cdot (Y + (X - Y)) + R_{2} \cdot (Y + (X - Y)) - \epsilon \cdot X\]
\[ R \cdot X > R_{1} \cdot Y + R_{2} \cdot (X - Y) + [(R_{1} \cdot (X - Y) + R_{2} \cdot Y) - \epsilon \cdot X]\]
Splitting reputation brings an influence benefit of:
\[
R_{1} \cdot (X - Y) + R_{2} \cdot Y - \epsilon \cdot X 
\]
We then examine the profitability condition for Sybil creation:
\[ R_1 \cdot (X - Y) + R_2 \cdot Y > \epsilon \cdot X \]
Given that $R_1$ and $R_2$ are fractions of $R$, the worst case happens when $R_1 = R_2 = R/2$. 
\[
\frac{R}{2} \cdot \frac{X}{2} + \frac{R}{2} \cdot \frac{X}{2} = \frac{R \cdot X}{2} > R_1 \cdot (X - Y) + R_2 \cdot Y  > \epsilon \cdot X
\]
Therefore, creating Sybil nodes is not profitable as soon as $R > 2 \cdot \epsilon$.
Similarly, the generalization of this demonstration to any number $K$ of Sybil nodes where each node loans an amount $X/K$ with a reputation $R/K$ yields:
\[K \cdot \frac{R}{K} \cdot \frac{X}{K} > \epsilon \cdot X \]
\[ R > K \cdot \epsilon\]
In practice, users would not accept an offline payment from a user who has a very low reputation. 

\subsection{Splitting a coin over Sybils}

Mitigating Sybil attacks in loan agreements involves addressing the case where Sybil nodes would attempt to loan the same coins multiple times. This deception would lead honest nodes to believe that they have access to more tokens than are available and would bias the offline confidence evaluation computations. This attack can be seen in Figure~\ref{fig:sc-av}. 

This attack is mitigated by the decentralized reputation scheme because a node generally splits its reputation over its potential Sybils, which means that Sybils are less likely to be chosen as transaction partners.
However, to fully counteract this attack, \name{} prevents nodes from accepting loans that involve tokens that have been used in established loans, which is achieved by its token-locking mechanism that locks the loan amount while a loan is active.

\subsection{Splitting a loan over Sybils}

\begin{figure*}[t]
    \centering
    \begin{subfigure}[b]{\columnwidth}
        \centering
        \includegraphics[width=\textwidth]{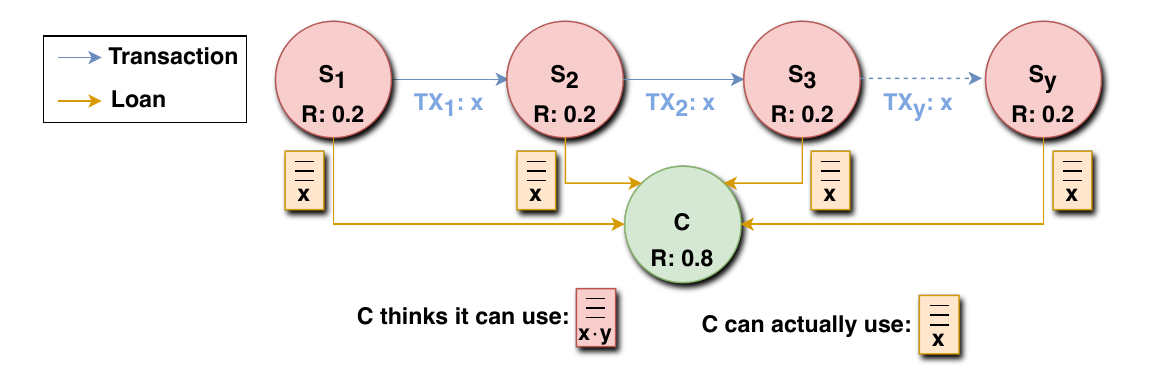}
        \caption{A Sybil attack utilizing multiple Sybil nodes and loan agreements to increase their maximum loan amount to abuse the incentives mechanism.}
        \label{fig:sc-av}
    \end{subfigure}%
    \vspace{0.3cm} 
    \begin{subfigure}[b]{\columnwidth}
        \centering
        \includegraphics[width=\textwidth]{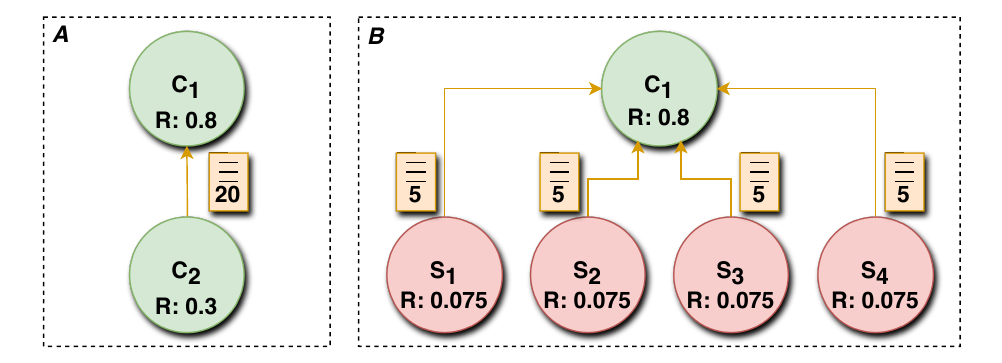}
        \caption{A loan agreement attack utilizing multiple Sybil nodes to loan fractions of another node's loan amount. The left part (A) shows a regular loan scenario. Part (B) shows the Sybil attack with nodes splitting to fractions of a loaning node's reputation.}
        \label{fig:sc-av2}
    \end{subfigure}
    \caption{Sybil attacks exploiting loan agreements in the network.}
\end{figure*}

Another possible loan agreement attack consists of creating multiple Sybil users that all loan tokens to the same user for a fraction of a more considerable transaction amount. 
This approach would reduce the risk of the recipient node, as apparently multiple nodes are available to contribute to the transaction amount, rather than relying solely on a single node for payment. Figure~\ref{fig:sc-av2} shows an example of two possible scenarios. The first scenario, $A$, shows two correct nodes $C_{1}$ and $C_{2}$ that are engaging in a loan agreement. For scenario $B$, the Sybil nodes vote for the same correct node for the same loan amount as $C_{2}$, with them having fractions of the amount and reputation. For the correct node, scenario $B$ is the most enticing as it will have to pay a lower interest rate and have more nodes that it can rely on, meaning less risk. 

However, this attack will not benefit Sybil nodes as the incentive mechanisms do not incentivize lower reputations. Furthermore, the influence of a node can be seen by its reputation times its loaned amount. Suppose the loaned amount is split among multiple Sybil nodes. In that case, the risk-reward benefit will only be applicable once the cumulative influence of the Sybil nodes is equal to that of the original loaning node. Reputation systems make these attacks difficult to execute because Sybil nodes are detected and penalized with a lower reputation.

\subsection{Preventing concurrent offline payments}

One might wonder whether \name{} can prevent a large number of offline payments from a given payer from happening concurrently in the network, which would lead the payer to realize a sizeable double-spending attack. To prevent this attack, a payment account should contain metadata that uniquely associates it to the identity of their owner, so that a payment receiver would only accept a payment if they can verify the physical presence of the payer. Such verification could be implemented using account metadata that can only be set when a payment account is created (i.e., the identity of the account owner), or using official documents.

\begin{figure*}[ht]
    \centering
    \begin{subfigure}{.45\textwidth} 
        \centering
        \includegraphics[width=\linewidth]{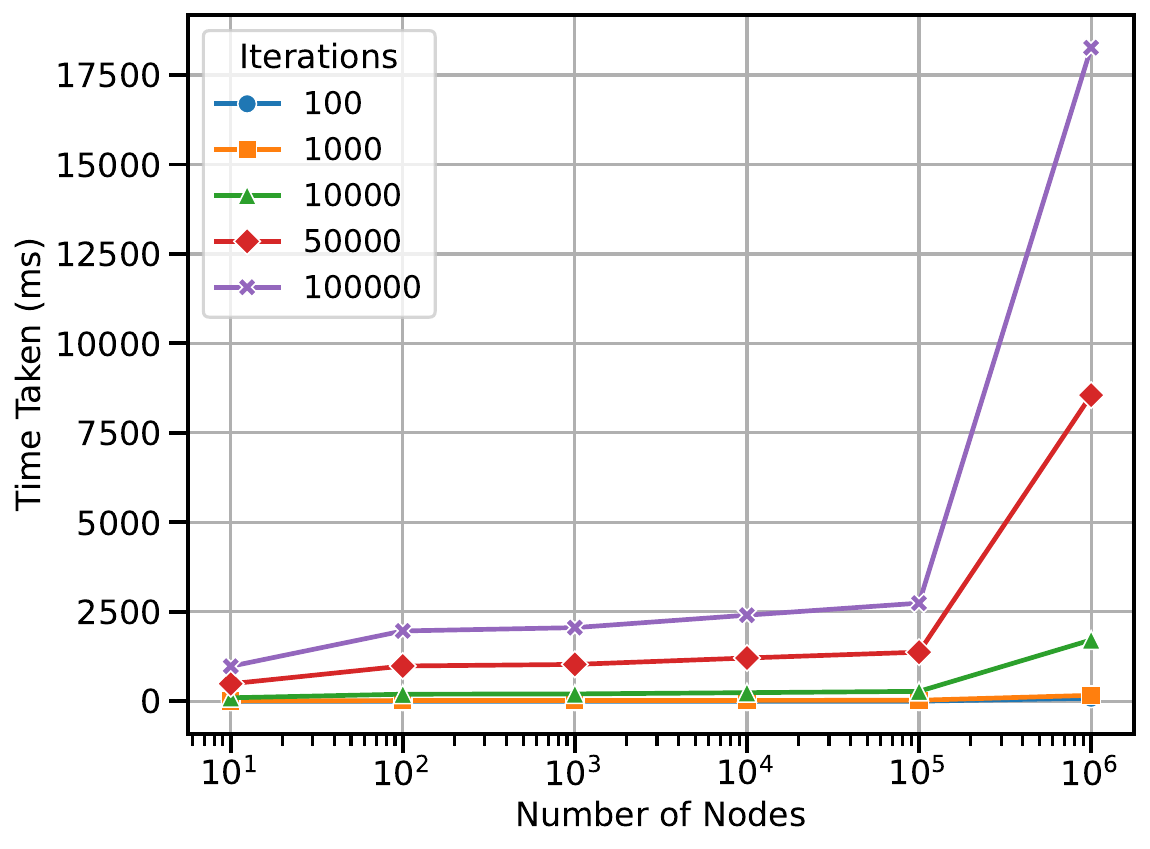}
        \caption{Without optimizations}
        \label{fig:performance_non_optimized}
    \end{subfigure}%
    \hfill
    \begin{subfigure}{.45\textwidth} 
        \centering
        \includegraphics[width=\linewidth]{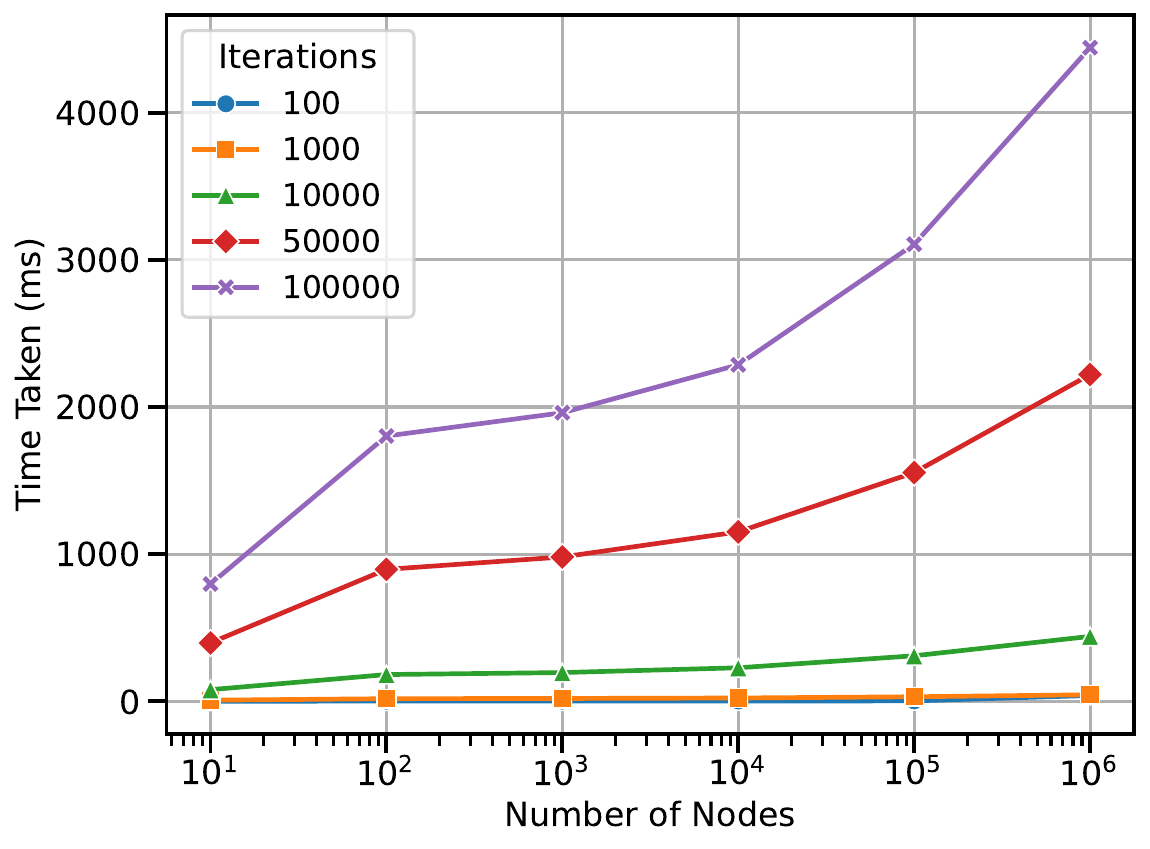}
        \caption{With optimizations}
        \label{fig:performance_optimized}
    \end{subfigure}
    \caption{Performance analysis of our algorithm without and with optimizations for the maximum amount loaned distribution.}
    \label{fig:performance_evaluation_plot_comparison}
\end{figure*}

\section{Performance Evaluation}
\label{sec:performance_evaluation}

In this section, we first describe our implementation of a prototype of \name{}. We then discuss the run time and accuracy of our offline confidence evaluation algorithm, the throughput and latency of the online part of \name{}, and the smart contract fees that offline transactions incur.

\subsection{Implementation details}

We developed \name{}'s smart contract for Ethereum~\cite{buterin2014next} using the Hardhat~\cite{hardhat} development environment and the Solidity~\cite{dannen2017introducing} programming language. We chose Ethereum for its widespread adoption, available testing tools, and extensive documentation.
Our smart contract encapsulates the logic for handling loan agreements, handling (offline) payments, and enforcing the rules governing the interactions between nodes within the network. 
We benchmarked \name{}'s smart contract on the Sepolia testnet, which replicates Ethereum's main network. 

\subsection{Confidence evaluation algorithm}

We evaluate the running time of our confidence evaluation algorithm by executing it on randomized graphs of $10$, $10^2$, $10^3$, $10^4$, $10^5$, and $10^6$ nodes, each connected by nine edges to other nodes, which mimics the average connectivity degree in the Lightning Network (LN)~\cite{martinazzi2020evolving}. We vary the number of randomized depth-first traversals between $10^2$ to $10^5$ intervals to study its impact.  
We use a $9$-hop maximum depth, reflecting the average LN hop distance, and a $0.95$ decay factor for node reputations. The loan capacity per node is randomly selected among the $0$, $10$, and $20$ values. Furthermore, the root node's reputation starts at $0.2$ reputation and is transacting $100$ tokens.

We provide a performance overview of the network's different iterations and node sizes in Table~\ref{tab:combined_performance_evaluation}. We evaluated our algorithm's performance with and without optimizations. In addition, we provide a graphical visualization of this evaluation, which can be found in Figure~\ref{fig:performance_optimized}. A high-reputation root node reduces the algorithm’s execution time, as it minimizes the need to traverse the network to locate additional lenders when nodes are unable to fulfill payments. We can see that after 50,000 iterations, the confidence interval width does not necessarily become much smaller than the 100,000 iterations. Thus, we can conclude that the accuracy of our algorithm is most optimal regarding computation time at around 50,000 iterations. We also compared the accuracy of our algorithm using a 95\% confidence interval in Table~\ref{tab:combined_ci_width_evaluation}. We noticed a significant increase in accuracy when comparing both tables, without and with optimizations, with little increase in accuracy when surpassing 50,000 iterations. 

\begin{figure*}[t]
    \centering
        \captionof{table}{Time (ms) necessary to retrieve the probability distribution regarding the maximum retrievable loan amount, comparing results without optimizations (Wo. Opt.) and with optimizations (With Opt.).} 
        \centering
        \begin{tabular}{l| c|c| c|c| c|c| c|c|}
         \textbf{Nodes} & \multicolumn{2}{c|}{\textbf{$10^2$ iter.}} & \multicolumn{2}{c|}{\textbf{$10^3$ iter.}} & \multicolumn{2}{c|}{\textbf{$10^4$ iter.}} & \multicolumn{2}{c|}{\textbf{$10^5$ iter.}} \\ \cline{2-9} 
         & \textbf{Wo. Opt.} & \textbf{With Opt.} & \textbf{Wo. Opt.} & \textbf{With Opt.} & \textbf{Wo. Opt.} & \textbf{With Opt.} & \textbf{Wo. Opt.} & \textbf{With Opt.} \\ \hline
        $10$ & 1.00 & 0.80 & 9.75 & 7.98 & 97.75 & 79.48 & 977.31 & 797.43 \\ \hline
        $10^2$ & 1.98 & 1.81 & 19.91 & 18.13 & 197.75 & 182.14 & 1965.24 & 1802.78 \\ \hline
        $10^3$ & 2.09 & 1.97 & 20.49 & 19.72 & 205.50 & 195.68 & 2057.84 & 1961.30 \\ \hline
        $10^4$ & 2.48 & 2.44 & 24.48 & 22.82 & 240.72 & 228.14 & 2405.89 & 2286.63 \\ \hline
        $10^5$ & 2.94 & 3.32 & 28.36 & 31.28 & 277.44 & 309.72 & 2742.24 & 3106.29 \\ \hline
        $10^6$ & 85.40 & 39.89 & 167.04 & 44.84 & 1713.39 & 441.44 & 18263.57 & 4441.81 \\ \hline
        \end{tabular}
        \label{tab:combined_performance_evaluation}
    
    \vspace{0.5cm} 
    
        \captionof{table}{Width of the 95\% confidence interval for the settled value of an offline payment of 100 coins without optimizations (Wo. Opt.) and with optimizations (With Opt.). For example, the top left value (9.47) indicates that the payment received online had a 95\% probability of being higher than $100-9.47=90.53$.}
        \centering
        \begin{tabular}{l|c|c|c|c|c|c|c|c|}
        \textbf{Nodes} & \multicolumn{2}{c|}{\textbf{$10^2$ iter.}} & \multicolumn{2}{c|}{\textbf{$10^3$ iter.}} & \multicolumn{2}{c|}{\textbf{$10^4$ iter.}} & \multicolumn{2}{c|}{\textbf{$10^5$ iter.}} \\ \cline{2-9} 
         & \textbf{Wo. Opt.} & \textbf{With Opt.} & \textbf{Wo. Opt.} & \textbf{With Opt.} & \textbf{Wo. Opt.} & \textbf{With Opt.} & \textbf{Wo. Opt.} & \textbf{With Opt.} \\ \hline
        $10$   & 9.47 & 6.99 & 3.46 & 2.26 & 1.08 & 0.75 & 0.34 & 0.23 \\ \hline
        $10^2$ & 16.35 & 13.82 & 5.05 & 4.30 & 1.56 & 1.38 & 0.50 & 0.44 \\ \hline
        $10^3$ & 11.74 & 11.04 & 4.10 & 3.76 & 1.34 & 1.20 & 0.42 & 0.38 \\ \hline
        $10^4$ & 18.53 & 13.94 & 5.96 & 4.60 & 1.85 & 1.47 & 0.58 & 0.47 \\ \hline
        $10^5$ & 22.27 & 24.31 & 7.51 & 6.72 & 2.32 & 2.13 & 0.75 & 0.68 \\ \hline
        $10^6$ & 24.74 & 22.39 & 96.17 & 7.80 & 29.42 & 2.45 & 13.20 & 0.78 \\ \hline
        \end{tabular}
        \label{tab:combined_ci_width_evaluation}
\end{figure*}

\begin{table*}[t]
\caption{Gas fee per operation on different blockchains (GWEI for ETH and Polygon, ALGO for Algorand). For Ethereum and Polygon, we use average fees on the network for low/medium/high load at the time of writing this paper.}
\centering
\begin{tabular}{l|c|c|c|c|c|c|c|}
\textbf{Operation} & \multicolumn{3}{c|}{\textbf{Ethereum (GWEI)}} & \multicolumn{1}{c|}{\textbf{Algorand (ALGO)}} & \multicolumn{3}{c|}{\textbf{Polygon (GWEI)}} \\ \cline{2-8}
 & \textbf{20} & \textbf{30} & \textbf{40} & \textbf{0.001} & \textbf{100} & \textbf{200} & \textbf{300} \\ \hline
 
Contract deployment 
& \$111.17 & \$166.76 & \$222.34 
& \$0.00017
& \$0.13 & \$0.26 & \$0.38 
\\ \hline

Offline payment \newline(good case) 
& \$2.21 & \$3.31 & \$4.41 
& \$0.00017
& \$0.0025 & \$0.0051 & \$0.0076 
\\ \hline

Offline payment \newline(avg. bad case) 
& \$5.90 & \$8.84 & \$11.79 
& \$0.00017
& \$0.0068 & \$0.014 & \$0.02 
\\ \hline

Creating a loan agreement 
& \$21.30 & \$32.09 & \$42.79 
& \$0.00017
& \$0.025 & \$0.049 & \$0.074 
\\ \hline

Closing a loan agreement 
& \$4.08 & \$6.12 & \$8.16 
& \$0.00017
& \$0.0047 & \$0.0094 & \$0.014 
\\ \hline

\end{tabular}
\label{table:smart_contract_costs_comparison}
\end{table*}

\begin{table*}[t]
\caption{Fees (in USD) associated with different quantities of loan agreements bundled by the smart contract on Ethereum and Polygon, with various network fees.}
\centering
\begin{tabular}{l|c|c|c|c|c|c|c|c|}

\textbf{\# Loan} & \multicolumn{4}{c|}{\textbf{Average Individual Fee}} & \multicolumn{4}{c|}{\textbf{Total Bundle Fee}} \\ \cline{2-9}
\textbf{Agreement}  & \multicolumn{2}{c|}{\textbf{Ethereum (GWEI)}} & \multicolumn{2}{c|}{\textbf{Polygon (GWEI)}} & \multicolumn{2}{c|}{\textbf{Ethereum (GWEI)}} & \multicolumn{2}{c|}{\textbf{Polygon (GWEI)}} \\ \cline{2-9}
\textbf{Created} & \textbf{20} & \textbf{30} & \textbf{100} & \textbf{200} & \textbf{20} & \textbf{30} & \textbf{100} & \textbf{200} \\ \hline

$10$
& \$20.27 & \$30.40 & \$0.0228 & \$0.0455
& \$202.68 & \$304.02 & \$0.23 & \$0.46 \\ \hline

$10^2$
& \$20.12 & \$30.17 & \$0.0226 & \$0.0452
& \$2,011.57 & \$3,017.35 & \$2.26 & \$4.52 \\ \hline

$10^3$
& \$20.12 & \$30.17 & \$0.0226 & \$0.0452
& \$20,115.68 & \$30,173.52 & \$22.60 & \$45.19 \\ \hline

\end{tabular}
\label{table:bundled_smart_contract_costs}
\end{table*}

\subsection{Throughput and Latency}
We measured the throughput and latency of \name{} using micro-benchmarks.

 We sent as many transactions as possible in one block to our smart contract. We measured the average block confirmation time of Sepolia's testnet, which is around 12 seconds, and the throughput of our smart contract. Afterward, we took Ethereum's gas block limit (30,000,000)~\cite{ethereumGas} and divided it by the average gas units used by the transactions multiplied by the average block confirmation time. 
We do not evaluate \name{}'s throughput and confirmation latency as they fundamentally depend on the blockchain on which its smart contract would be deployed, or on the layer-2 (L2) solutions that could be used. Additionally, the online settlement of offline transactions is currently executed in \name{}'s smart contract (i.e., exploring the loan network), which is a task that could be executed by users directly to improve performance.  

\subsection{Transaction Fees}
Executing an individual procedure of the smart contract incurs fees. 
\name{}'s smart contract fees are shown in Table~\ref{table:smart_contract_costs_comparison}.
The costs are shown in USD at the time of April $16^{th}$, 2024: 3,030 USD per ETH (Ethereum), 0.17 USD per ALGO (Algorand), and 0.69 USD per MATIC (Polygon).
This table displays the smart contract operation costs when deployed on different L1 blockchains, such as Ethereum or Algorand, or an L2 solution, such as Polygon. We take reasonable average fee prices for each network to gain insight into the possible costs for each operation executed on the smart contract.
Bundling transactions on the smart contract would make it possible to decrease fees further. Table~\ref{table:bundled_smart_contract_costs} shows the fees associated with different bundled loan agreement creations. We compare Ethereum, as our smart contract was already implemented in Solidity, with Polygon, as it is Ethereum Virtual Machine (EVM) compatible, meaning the smart contract is easily deployed on the Polygon blockchain. 
Bundling agreement creations would decrease these fees further. 

\section{Conclusion}
\label{sec:conclusion}

We presented a novel framework for reputation-weighted loans that facilitates secure offline transactions by integrating network-based lending with blockchain technology. Our system, \name{}, leverages trust and reputation to tolerate and minimize risks such as double-spending and enhancing security in decentralized networks. By implementing a smart contract, an effective loan management algorithm, and an incentive model, we demonstrated that \name{} supports scalable and efficient offline payments with minimal computational demands. Future work will focus on advancing privacy measures, mitigating contagion risks, and improving the incentive formula. This research is pivotal in bridging conventional financial systems and the digital economy, highlighting \name{}'s potential to revolutionize offline payments.
 
Possible optimizations of \name{} include compressing smart contract fields, integrating it with another L1 or L2 solution to reduce costs and increase performance, and using multiple smart contracts that would manage subsets of users (i.e., sharding). 

Furthermore, it is a promising direction to explore how to leverage privacy-preserving technologies, e.g., Zero-Knowledge Proofs (ZKPs), to ensure transaction confidentiality without compromising transparency and verifiability.

\printbibliography

\appendices
\section{Smart Contract Implementation Details} \label{app:routines}
This appendix details the core procedures implemented by \name{}'s smart contract. These routines handle the creation of loan agreements, utilization of loaned tokens in case of shortfalls, and token transfers between users.

\subsection{Variables and Data Structures}
We summarize below the key variables and mappings used in the smart contract routines:

\begin{itemize}
    \item \textbf{Participants}: $\mathcal{A}$ (lender), $\mathcal{B}$ (borrower)
    \item \textbf{Public Keys}: $\text{pk}\mathcal{A}$, $\text{pk}\mathcal{B}$
    \item \textbf{Loan Amount}: $L \in \mathbb{R}^+$
    \item \textbf{Repayment Time}: $T_r \in \mathbb{T}$
    \item \textbf{Dispute Resolution Mechanism}: $D_r$
    \item \textbf{Agreement Duration}: $T_d \in \mathbb{T}$
    \item \textbf{Minimum Open Time}: $T_o \in \mathbb{T}$
    \item \textbf{Opening Fee}: $F_o \in \mathbb{R}^+$
    \item \textbf{Agreement ID}: $agreementId \in \mathbb{N}$
    \item \textbf{Balance Mapping}: $\text{balances}[\cdot]$
    \item \textbf{Agreements Mapping}: $\text{agreements}[\cdot]$
    \item \textbf{Loaned Amounts Mapping}: $\text{loanedAmounts}[\cdot][\cdot]$
    \item \textbf{Last Loan Agreement ID Mapping}: $\text{lastLoanAgreementId}[\cdot][\cdot]$
\end{itemize}

\subsection{Core Procedures}
\begin{protocol}{InitiateLoan} \label{alg:initiateLoan}
\textbf{Data:} Lender $\mathcal{A}$, Borrower $\mathcal{B}$, Loan Amount $L$, Repayment Time $T_r$, Dispute Resolution Mechanism $D_r$, Agreement Duration $T_d$, Minimum Open Time $T_o$, Opening Fee $F_o$.

\sbline

\textbf{Result:} Loan agreement created, tokens transferred, and state updated accordingly.

\sbline

\begin{enumerate}
    \item \textbf{Input Validation:}
    \begin{enumerate}
        \item Ensure that $\mathcal{A} \neq \mathcal{B}$.
        \item Validate that all monetary values ($L$, $F_o$) are positive.
        \item Confirm that time parameters ($T_r$, $T_d$, $T_o$) are within acceptable ranges.
    \end{enumerate}
    \item \textbf{Balance Check:}
    \begin{enumerate}
        \item Compute $totalDebit \gets L + F_o$.
        \item \textbf{If} $\text{balances}[\mathcal{A}] < totalDebit$ \textbf{then}
        \begin{enumerate}
            \item Abort with error: Insufficient funds in $\mathcal{A}$'s account.
        \end{enumerate}
    \end{enumerate}
    \item \textbf{Agreement ID Generation:}
    \begin{enumerate}
        \item $agreementId \gets \text{generateUniqueAgreementId}()$.
    \end{enumerate}
    \item \textbf{Create Agreement Record:}
    \begin{enumerate}
        \item $\text{agreements}[agreementId] \gets \{\mathcal{A}, \mathcal{B}, L, T_r, D_r, T_d, T_o, F_o\}$.
    \end{enumerate}
    \item \textbf{Update Mappings:}
    \begin{enumerate}
        \item $\text{lastLoanAgreementId}[\mathcal{A}][\mathcal{B}] \gets agreementId$.
        \item $\text{loanedAmounts}[\mathcal{A}][\mathcal{B}] \gets \text{loanedAmounts}[\mathcal{A}][\mathcal{B}] + L$.
    \end{enumerate}
    \item \textbf{Transfer Tokens:}
    \begin{enumerate}
        \item \textbf{Debit Lender:}
        \begin{enumerate}
            \item $\text{balances}[\mathcal{A}] \gets \text{balances}[\mathcal{A}] - totalDebit$.
        \end{enumerate}
        \item \textbf{Credit Borrower:}
        \begin{enumerate}
            \item $\text{balances}[\mathcal{B}] \gets \text{balances}[\mathcal{B}] + L$.
        \end{enumerate}
    \end{enumerate}
    \item \textbf{Emit Events:}
    \begin{enumerate}
        \item Emit \texttt{AgreementCreated(agreementId, $\mathcal{A}$, $\mathcal{B}$, $L$)}.
        \item Emit \texttt{TokensTransferred($\mathcal{A}$, $\mathcal{B}$, $L$)}.
    \end{enumerate}
\end{enumerate}
\end{protocol}

\begin{protocol}{UseLoanedTokens} \label{alg:useLoanedTokens}
\textbf{Data:} Borrower $\mathcal{B}$, Required Amount $A_r$.

\sbline

\textbf{Result:} Loaned tokens are utilized to cover the shortfall.

\sbline

\begin{enumerate}
    \item \textbf{Initialize Variables:}
    \begin{enumerate}
        \item Set $loanedAmount \gets 0$.
        \item Set $remainingAmount \gets A_r$.
    \end{enumerate}
    \item \textbf{Iterate Over Lenders:}
    \begin{enumerate}
        \item \textbf{For each} Lender $\mathcal{L}$ in $\text{lendersOf}[\mathcal{B}]$ \textbf{do}:
        \begin{enumerate}
            \item Retrieve $availableLoan \gets \text{loanedAmounts}[\mathcal{L}][\mathcal{B}]$.
            \item \textbf{If} $availableLoan > 0$ \textbf{then}:
            \begin{itemize}
                \item Compute $usedAmount \gets \min(remainingAmount, availableLoan)$.
                \item Update $\text{loanedAmounts}[\mathcal{L}][\mathcal{B}] \gets \text{loanedAmounts}[\mathcal{L}][\mathcal{B}] - usedAmount$.
                \item Update $\text{balances}[\mathcal{B}] \gets \text{balances}[\mathcal{B}] + usedAmount$.
                \item Update $loanedAmount \gets loanedAmount + usedAmount$.
                \item Update $remainingAmount \gets remainingAmount - usedAmount$.
                \item \textbf{If} $remainingAmount = 0$ \textbf{then}:
                \begin{itemize}
                    \item \textbf{Break} out of the loop.
                \end{itemize}
                \item \textbf{End If}
            \end{itemize}
            \item \textbf{End If}
        \end{enumerate}
    \end{enumerate}
    \item \textbf{Check Loaned Amount:}
    \begin{enumerate}
        \item \textbf{If} $loanedAmount < A_r$ \textbf{then}:
        \begin{itemize}
            \item Abort with error: Not enough loaned tokens available.
        \end{itemize}
    \end{enumerate}
    \item \textbf{Emit Event:}
    \begin{enumerate}
        \item Emit \texttt{LoanTokensUsed($\mathcal{B}$, $A_r$)}.
    \end{enumerate}
\end{enumerate}
\end{protocol}

\begin{protocol}{TransferTokens} \label{alg:transferTokens}
\textbf{Data:} Sender $\mathcal{S}$, Recipient $\mathcal{R}$, Transfer Amount $A_t$.

\sbline

\textbf{Result:} Tokens transferred from $\mathcal{S}$ to $\mathcal{R}$, utilizing loaned tokens if necessary.

\sbline

\begin{enumerate}
    \item \textbf{Compute Available Balance:}
    \begin{enumerate}
        \item $availableBalance \gets \text{balances}[\mathcal{S}]$.
    \end{enumerate}
    \item \textbf{Check for Sufficient Balance:}
    \begin{enumerate}
        \item \textbf{If} $availableBalance \geq A_t$ \textbf{then}
        \begin{enumerate}
            \item Proceed to Step 3.
        \end{enumerate}
        \item \textbf{Else}
        \begin{enumerate}
            \item $shortfall \gets A_t - availableBalance$.
            \item \textbf{Invoke} \texttt{UseLoanedTokens}($\mathcal{S}$, $shortfall$).
            \item Recompute $availableBalance \gets \text{balances}[\mathcal{S}]$.
            \item \textbf{If} $availableBalance < A_t$ \textbf{then}
            \begin{enumerate}
                \item Abort with error: Insufficient funds after utilizing loaned tokens.
            \end{enumerate}
        \end{enumerate}
    \end{enumerate}
    \item \textbf{Update Balances:}
    \begin{enumerate}
        \item $\text{balances}[\mathcal{S}] \gets \text{balances}[\mathcal{S}] - A_t$.
        \item $\text{balances}[\mathcal{R}] \gets \text{balances}[\mathcal{R}] + A_t$.
    \end{enumerate}
    \item \textbf{Emit Event:}
    \begin{enumerate}
        \item Emit \texttt{TokensTransferred($\mathcal{S}$, $\mathcal{R}$, $A_t$)}.
    \end{enumerate}
\end{enumerate}
\end{protocol}

\clearpage
\section{Illustrative Example of Payment Protocol}
\label{app:paymentprotocol}

This appendix presents a step-by-step visual illustration of a typical offline payment processed through \name{}'s smart contract. The scenario shows a user, Bob, attempting to pay another user, Charlie, while offline, using a loan agreement previously set up with Alice. The diagram outlines both the successful and unsuccessful paths depending on Bob's behavior, and how the system resolves the transaction upon reconnecting online.

Each box represents the state of a user's tokens ($T$) and reputation ($R$) at a particular moment. Solid arrows depict token flows and reputation changes, while dashed or labeled arrows indicate smart contract intervention or fallback resolution mechanisms. This example complements the pseudocode routines described in Appendix~\ref{app:routines} and the design outlined in Section~\ref{sec:smartcontract}.

\begin{strip}
    \centering
    \includegraphics[width=0.95\textwidth]{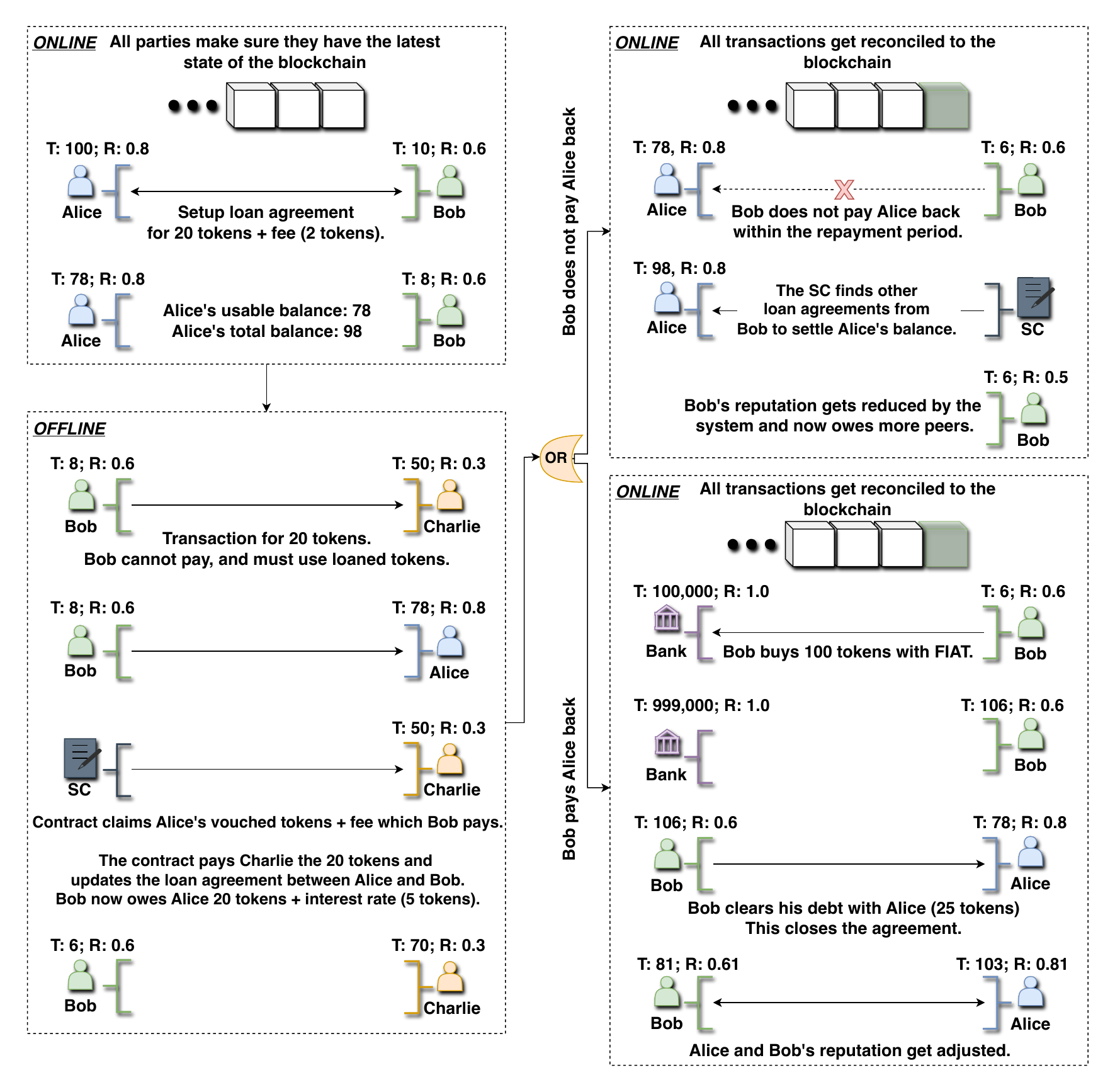}
    \captionof{figure}{An overview of \name{}'s payment protocol displaying loan agreement creation, an offline payment utilizing the loan, and the repayment possibilities.}
    \label{fig:payment-protocol}
\end{strip}

\end{document}